\documentclass[11pt,a4paper]{article}
\usepackage[T1]{fontenc}
\usepackage[utf8]{inputenc}
\usepackage[margin=2.5cm]{geometry}
\usepackage{amsmath,amssymb}
\usepackage{booktabs}
\usepackage{graphicx}
\usepackage{natbib}
\usepackage[hidelinks]{hyperref}
\hypersetup{pdftitle={Ranking the wrong places: vulnerability assessment, flood losses and risk sharing in Italy and Europe}, pdfauthor={Stefano Blando}}
\usepackage{setspace}
\usepackage{mathpazo}
\usepackage{microtype}
\usepackage{caption}
\usepackage{titlesec}
\titleformat{\section}{\normalfont\large\bfseries}{\thesection.}{0.8em}{}
\titleformat{\subsection}{\normalfont\normalsize\bfseries}{\thesubsection.}{0.8em}{}

\title{\LARGE\bfseries Ranking the wrong places: vulnerability assessment, flood losses
and risk sharing in Italy and Europe}

\author{
  \textbf{Stefano Blando}\\[0.35em]
  \small L'EMbeDS Department of Excellence, Scuola Superiore Sant'Anna, Pisa\\[0.1em]
  \small Department of Computer Science, University of Pisa\\[0.25em]
  \small\href{mailto:stefano.blando@santannapisa.it}{\texttt{stefano.blando@santannapisa.it}}
}
\date{\small September 2026}

\begin{document}
\maketitle

\vspace{-0.5em}
\begin{abstract}
\noindent\small
Before a flood, governments decide where to invest in protection; after it, how much of the loss
to compensate. The European Union informs the first decision with a composite index that ranks
every subnational unit by vulnerability. This paper tests the ranking by comparing Italian
provinces struck by the same flood between 2005 and 2024. The published index does
not order provinces by the losses they suffer, with or without controls for hazard, exposure and income
($\rho$ between $-0.06$ and $0.09$). The failure lies in what it weights. As published, the index mostly measures economic
development, while losses follow water: they arrive in recurring groups of provinces that share a
river basin, and more than half of the ranking error lies between those groups. In a pre-registered test that rebuilds it
and eight rivals at every forecast origin from 2010 to 2020, the index comes last on both questions
an allocator asks. On where damage will land, its top ten provinces capture 16 per cent of later
damage, less than a random draw; Italy's public flood hazard map, blind to losses,
captures 44. On who loses more from the same water, its rank correlation with later losses is
0.03; its own components, reweighted on past losses, reach 0.42. Neither repair needs new data. The
misranked loss is paid locally. Of EUR 12.3 billion recorded over 2013--2022, 85 per cent was not
covered by national transfers, and prevention funding, scored the same way, does no better than
chance. Because correlated losses cluster within borders, national
pools in Europe need 39 per cent more reserve than cross-border pools of the same size. The index fails the same way across 1{,}142
European units, where the repairs are weaker.

\end{abstract}
\vspace{1em}

\section{Introduction}

A government facing floods makes two decisions with public money: where to protect in advance, and
how much of the loss to compensate afterwards. The political economy of that pair is well
documented. \citet{healy2009} show that voters reward relief spending and not preparedness, and
estimate that one dollar of preparedness is worth about fifteen dollars of damage avoided. This
paper asks the next question: when preparedness money is spent, does the information used to
direct it point to the places that lose?

In the European Union that information has a name. For every NUTS-3 unit the Risk Data Hub of the
Joint Research Centre publishes a Local Vulnerability Index, built from forty-nine socio-economic,
physical and institutional components and meant to compare territories before an event. An index
of this kind is a claim about the future: the places it ranks as fragile are the places that will
lose most when the water comes. If the claim fails, prevention guided by it goes to the wrong
places, and the losses it misses fall on whoever lacks cover.

The economics of disasters has mostly measured what happens after the event: growth
\citep{felbermayr2014}, aid \citep{stromberg2007}, local institutions \citep{barone2014}, political
responses \citep{cerqua2023} and the regional consequences of wildfires and earthquakes
\citep{meier2023,aksoy2025}. Flood research
compares damage models with one another \citep{jongman2012} and debates risk management under
unprecedented events \citep{kreibich2022}, but rarely scores the instrument that ranks places
before the water arrives against what the water does. Vulnerability indices have been validated
against outcomes at country scale \citep{birkmann2022,formetta2019}. The closest European use of
the index tested here is \citet{ronco2026}, who enter it, averaged over time, as one feature of a
pooled model of regional flood losses, but do not score it as an ordering of the units a flood
hits. In the United States, \citet{rufat2019} find that four social vulnerability models explain
property loss after Hurricane Sandy poorly, and \citet{tellman2020} that individual indicators
predict flood deaths and damage better than the composite does. Other work assesses such indices
by internal sensitivity and construction criteria \citep{tate2012,spielman2020}. That a composite
can miss therefore needs no new demonstration. What remains open is why it misses, whether the
index the European Union publishes misses at the unit at which money moves, and who pays when it
does.

The test compares provinces hit by the same flood of 2005--2024 and treats what
remains in the loss equation, after hazard, exposure and income, as revealed vulnerability. The
published index is uncorrelated with it ($\rho = 0.078$), and stays so when any of those controls
is dropped. The index's dominant
factor is socio-economic and blind to loss, while the factor that tracks loss, the national
hydraulic hazard maps, barely enters it. Losses arrive not along any single ordering of provinces
but in recurrent groups that sit inside river basins and hold more than half the variance of the
index's error. Rebuilt at every forecast origin from 2010 to 2020, nine rankings separate two
questions: where damage will land, and who loses more when the same water arrives. The index comes
last on both. The national flood hazard map, public and blind to losses, answers the first best;
the index's own components, reweighted on past losses, answer the second. The loss the index
misses is paid locally: national transfers cover about 15 per cent of recorded damage, and EUR
21.50 billion of prevention funding follows territory and the generation of water rather than
damage.

The paper makes three contributions, and treats the index throughout as a forecasting model would
be treated: scored out of time, against simple baselines, with the decisive comparisons registered
before they were run. It scores the European index, as a ranking, against losses
within events, in Italy and on 1{,}142 European units, and traces the failure to the weights. It
shows, at every forecast origin, that information already public or already collected outranks
it, while the prevention money actually spent, scored on the same yardstick, does no better than
chance. And it measures
the dependence of losses across units, the systemic side of flood risk, on recorded subnational
data. That floods strike many places at once
and that national pooling is costly is known \citep{berghuijs2019,jongman2014,prettenthaler2017}.
What is added here is the measurement on recorded losses, its link to the ranking error, and the
reserve cost of pooling along administrative lines.

\section{Data}

\paragraph{Losses.} Event-level impacts come from the Risk Data Hub, which reports casualties,
affected population and economic loss by event and NUTS-3 unit. The Italian sample contains 364
events between 1980 and 2025, 306 of them hydro-meteorological. Of these, 175 events on 107 provinces are
recorded in at least two provinces and enter the structural analysis. The flood records derive
from the HANZE compilation \citep{paprotny2018hanze,paprotny2024}, whose known under-reporting is
treated as a floor rather than corrected \citep{paprotny2018trends}.

\paragraph{Hazard and exposure.} Precipitation and runoff come from ERA5-Land
\citep{munozsabater2021} over each event window and over the same calendar window in every year
from 1950 to 2025, so each province-event carries a hazard percentile relative to its own
climatology. Built surface, gridded population and the share of territory at low elevation above
the drainage network come from the Global Human Settlement Layer and MERIT Hydro.

\paragraph{Assessed vulnerability.} The Local Vulnerability Index, its four dimensions and its
forty-nine components are taken from the Risk Data Hub for 2005--2025, with a matching panel for
1{,}203 European units.

\paragraph{Transfers and insurance.} Public transfers come from the national emergency dataset of
\citet{gatto2023}: funds approved by central government for 123 emergencies over 2013--2022,
allocated to provinces by the number of municipalities under declaration \citep[see
also][]{clo2025}. Private insurance density is the provincial distribution of non-life premiums
net of motor, published by IVASS.

\paragraph{Prevention expenditure.} The ReNDiS register of the national environmental protection
agency lists 28{,}278 georeferenced soil-defence works, 28{,}085 of which fall inside a provincial
polygon. Funding totals EUR 21.50 billion, of which EUR 8.01 billion is concluded and EUR 10.04
billion has neither been concluded nor entered execution. Project dates, absent from the register,
come from the dataset of \citet{ricciotti2024}, which links it to the treasury and cohesion-policy
platforms (Section~\ref{sec:prevention}).

\paragraph{Hydrography.} The water graph is built on the 2{,}275 HydroBASINS level-9 units covering
Italy \citep{lehner2013}. Runoff generated in each basin during each event window is routed
downstream and credited to every province the path reaches. For the median province, 68 per cent
of the water arriving in a flood window is generated outside its boundary.

\section{Research design}

For province $i$ in event $e$,
\begin{equation}
y_{ie} = \alpha_e + \beta\,\mathrm{VI}_{it} + \gamma' x_{ie} + \varepsilon_{ie},
\end{equation}
where $y_{ie}$ is log economic loss over provincial GDP, $\alpha_e$ an event fixed effect,
$\mathrm{VI}_{it}$ the published index in the year of the event, and $x_{ie}$ collects hazard
percentile, built density, population and lagged income per head. The event effect absorbs the
storm's severity and the reporting conventions of its year, so identification comes from provinces
hit by the same event. The index exists from 2005, which leaves 317 province-events in 66 events on
96 provinces. The median province is observed three times.

A province's revealed vulnerability is the mean residual of this equation across its events: a
quantity estimated from losses, against which the index is scored. It is not noise. Splitting each
province's events into halves gives a split-half correlation of $0.375$ in Italy, $0.534$ in Europe
and $0.497$ in Europe excluding Italy, and the revealed trait is orthogonal to the published index
in all three samples (correlations between $0.01$ and $0.03$).

Conley standard errors with a 200 km cutoff are preferred throughout, since neighbouring provinces
receive the same water \citep{conley1999}, with clustering by unit reported beside them. A wild
cluster bootstrap on macro-regions gives $p = 0.033$ in Italy, on five clusters, too few to rely on, and
$p = 0.305$ in Europe, on twenty-five. Randomisation inference, permuting the index across the units hit by the
same event, rejects the sharp null that the index carries no information about which unit loses
more ($p < 0.001$ in all three samples). Lee bounds on selection into having a recorded figure
\citep{lee2009} exclude zero in Italy and include it in Europe, and the Oster $\delta$ is negative
throughout \citep{oster2019}.

\section{Ranking performance of the a priori index}
\label{sec:index}

\subsection{What the index contains}

Nineteen of the forty-nine components of the Italian index are a national figure repeated for every
province. Across Europe the median share of such components is 39.2 per cent, and Italy, at 38.8,
is typical. Within a country, two fifths of the index cannot distinguish anything.

A parallel analysis retains six factors. The first is socio-economic and correlates at $0.818$ with
the published index and at $0.035$ with the revealed trait. The second loads on the national
hydraulic hazard maps and correlates at $0.368$ with the revealed trait and at $0.011$ with the
index. A factor blind to loss dominates the index; the factor that tracks loss barely enters it.

\subsection{The ranking}

As published, the index shows no correlation with the revealed trait in Italy and a weakly negative
one in the two European samples (Table~\ref{tab:rank}). Its own components, re-aggregated with
weights fitted out of fold so that no province contributes to the weights that rank it, reach
$0.487$, $0.315$ and $0.212$, and the paired bootstrap difference excludes zero in all three
samples. The gap does not come from thinly observed provinces. Replacing each province's mean
residual with its empirical-Bayes estimate, which shrinks a province seen once or twice toward the
mean, the index scores $0.107$, $-0.019$ and $-0.004$ and the re-aggregation $0.476$, $0.327$ and
$0.228$, with paired differences of $0.370$ $[0.072, 0.667]$, $0.346$ $[0.208, 0.475]$ and $0.232$
$[0.117, 0.351]$. Only among the thirty-six Italian provinces seen four times or more does the gap
narrow, to $0.138$.

The failure does not come from the controls. Comparing only provinces struck by the same event,
with no other adjustment, the index scores $-0.013$, and across thirty-six specifications that drop
income, exposure or hazard, change the hazard measure or the minimum event and province size, it
stays between $-0.06$ and $0.09$. Without the event effect it reaches $0.25$, but that correlation
runs between events: high-index provinces are hit by storms with larger average losses ($0.31$), not
harder than their neighbours by the same storm.

By Spearman--Brown, a split-half correlation of $0.375$ implies a ceiling of $0.74$ on the
correlation any ranking can reach with the revealed trait. Re-aggregation reaches two thirds of that
ceiling; the published index, a tenth. Section~\ref{sec:tournament} repeats the comparison out of
time, at every forecast origin.

\begin{table}[t]
\centering\small
\caption{Rank correlation with revealed vulnerability.}
\label{tab:rank}
\begin{tabular}{lccc}
\toprule
& Italy & Europe & Europe excl.\ Italy \\
\midrule
Published index & $0.078$ & $-0.038$ & $-0.012$ \\
Components re-aggregated out of fold & $0.487$ & $0.315$ & $0.212$ \\
Paired difference & $0.409$ & $0.354$ & $0.224$ \\
\quad bootstrap interval & $[0.091,\,0.698]$ & $[0.216,\,0.491]$ & $[0.097,\,0.340]$ \\
\bottomrule
\end{tabular}
\end{table}

Inside the damage equation the index does carry information ($0.52$, $t = 2.9$, in Italy), but
mostly through its movement over time within a province ($1.25$, $t = 2.9$) rather than its level
($0.24$, $t = 1.85$). It says more about when a province is fragile than about which province is,
and only the second can order provinces for an allocation.

\begin{figure}[t]
\centering
\includegraphics[width=\textwidth]{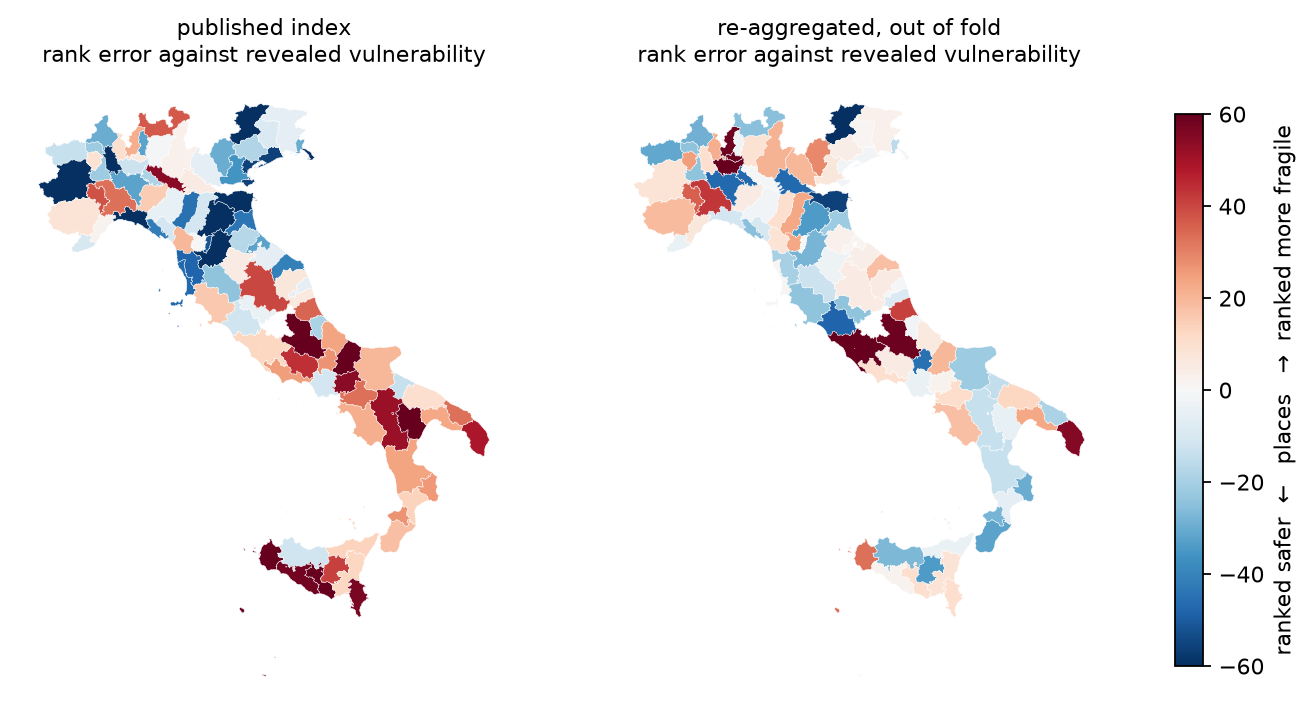}
\caption{Rank error against revealed vulnerability, Italian provinces. Red is ranked more fragile
than losses warrant, blue ranked safer. Left: the published index. Right: its own components
re-aggregated on losses, with the weights of each province estimated without that province.}
\label{fig:map}
\end{figure}

Across space, the errors have a geography. Florence, Trieste, Bologna, Genoa and Turin are placed far safer than
their losses warrant, L'Aquila, Trapani and Matera far more fragile (Figure~\ref{fig:map}): the
error maps development. Re-aggregation removes the gradient rather than reversing it. The mean rank
error falls from 30.2 to 20.5 places out of ninety-six, and what remains is scattered.

\subsection{What the misranking costs}

\begin{figure}[t]
\centering
\includegraphics[width=0.92\textwidth]{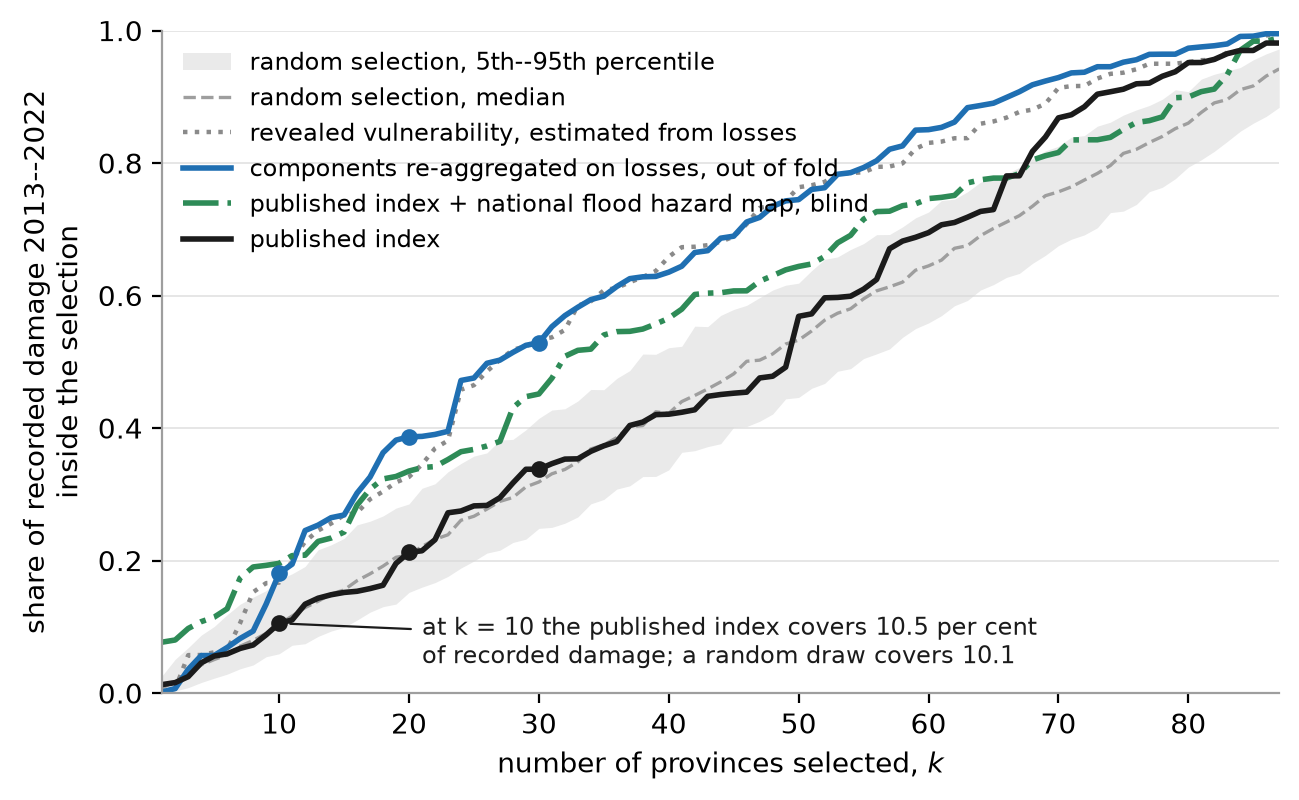}
\caption{Share of recorded damage 2013--2022, in 2022 prices, falling inside the top $k$ Italian
provinces of each ranking. The dash-dotted line averages the published index and the national
flood hazard map without reference to any loss (Section~\ref{sec:map}). The band is the 5th to
95th percentile of four hundred random selections of size $k$. Markers are the three cut-offs
carried through the bootstrap.}
\label{fig:capture}
\end{figure}

A rank correlation is not a policy quantity. The index exists to order places for a budget that
cannot cover all of them, so the measure with a unit attached is coverage: how much of the damage
that occurred fell inside the $k$ provinces the index would have selected?
Figure~\ref{fig:capture} reports that curve for 2013--2022, in 2022 prices, beside random
selections of the same size.

Selection by the published index resembles a lottery. Its top ten, twenty and thirty provinces
cover 10.5, 21.2 and 33.8 per cent of recorded damage, against 10.1, 21.2 and 32.0 for the median
random draw. On damage recorded after 2014, the window an allocator would have faced, it falls
below chance: 5.4 per cent against 12.2 at $k = 10$. Ranking by population covers 8.7 per cent, so
the curve is not a size effect. The re-aggregated ranking covers 18.1, 38.7 and 52.9 per cent, with
probabilities of a non-positive gain of $0.060$, $0.045$ and $0.040$ when events are resampled and
the weights refitted in each replicate: EUR 1.34 to 2.42 billion of damage brought inside the
selection. These are shares of recorded damage, not damage prevented. The gain is Italian (in
Europe the same probabilities are $0.155$, $0.180$ and $0.090$) and in euros rather than burden:
per inhabitant it shrinks to $0.044$ at $k = 10$ ($P = 0.27$).

\subsection{Where the signal is}
\label{sec:map}

The re-aggregated ranking is supervised on losses and the published one is not. Two rankings built
from the same components without any loss figure separate the two effects: equal weights on the
components that vary within the country, signed by their correlation with the index, and the first
principal component with the same sign. Both rank worse than the published index, by $0.144$ and
$0.106$, in every one of two hundred bootstrap replicates, and on coverage at $k = 10$ all three are
indistinguishable from a random draw. The Joint Research Centre weighting carries information of
its own, and no weighting of the components that ignores losses contains the repair.

Other inputs, assessed before any event, do. Of eighty-four a-priori measures screened against
the revealed trait with the false discovery rate controlled (index components, well-being
indicators of the national statistical institute, and the layers of the national hydraulic and
landslide hazard maps \citep{iadanza2021}), the hazard maps win. Firms in the
high-hazard flood band correlate with the revealed trait at $0.355$ ($q = 0.007$), buildings at
$0.346$ and residents at $0.333$, against $0.078$ for the index. They do not proxy exposure (the
largest correlation with any control is $0.148$), and the index does not contain them (correlation
$-0.077$). Entered together in the within-event equation, the index keeps $0.248$ ($t = 2.99$ under
Conley errors) and the map takes $0.193$ ($t = 3.66$). A falsification separates the map's layers:
within floods, the hydraulic layer predicts loss at $+0.0040$ ($t = 2.44$) and the landslide layer
at $-0.0036$ ($t = -2.03$), a sign change a generic ``physical'' proxy would not produce.

Averaging the index with the three high-hazard layers at equal weight, with no reference to any
loss, raises the rank correlation from $0.078$ to $0.322$, a gain positive in all two hundred
bootstrap replicates. On damage recorded after 2014 it covers 23.0 per cent at $k = 10$, against
5.4 for the index and 12.2 for the median random draw ($P = 0.035$): EUR 3.19 billion of damage the
index left outside the selection. Unlike the supervised gain, it survives per inhabitant ($0.099$,
$P = 0.10$).

At this split the result is not look-ahead. The 2021 map post-dates part of the scoring window, and hazard plans
are revised after floods. Yet a pre-registered rerun on the 2017 map \citep{ispra2018} covers 24.1 per
cent of post-2014 damage at $k = 10$, against 5.5 for the index ($P = 0.013$), though revisions made
in 2015--2016 cannot be excluded. It does depend on the map being national. With the one pan-European
alternative, the GloFAS hazard layer simulated for large rivers, the Italian combination covers 9.3
per cent of the top decile's damage against 8.9 for a random draw and 23.0 with the national map,
and in a pre-registered test on sixteen other countries it adds nothing to the index. France's
national flood map ranks French revealed vulnerability at $0.43$ against $0.01$ for the index, but
its capture gain at $k = 10$ is not significant ($P = 0.11$). The Italian map is public and
maintained by the environmental agency: improving the index requires using what the same
administration publishes.

\subsection{Two questions, two rankings}
\label{sec:tournament}

Every comparison so far rests on one split in time, and an allocator asks two questions: where
damage will land, and who loses more when the same water arrives. In a pre-registered tournament,
nine rankings are rebuilt at each forecast origin from 2010 to 2020 with what an allocator had at
that date: the index and components published that year, the events up to it, population and the
hazard map. The first question is scored by capture of the damage recorded after the origin at
$k = 10$; the second by the rank correlation with the revealed trait of later events, which holds
hazard fixed and is estimable at origins 2010--2014. Intervals come from a block bootstrap over
events, with every supervised ranking refitted in each replicate (Table~\ref{tab:tournament}).

The published index comes last on both. Its top ten provinces capture 15.9 per cent of later damage
on average, less than a random draw at nine of eleven origins, and its rank correlation with later
revealed vulnerability averages $0.03$.

A different ranking wins each question. On where damage lands, the national flood hazard map alone
captures 43.9 per cent, is first at nine origins and ahead of the index at all eleven ($P = 0.04$).
On who loses more, the components reweighted on past losses reach $0.42$, ahead of the index at
all five eligible origins ($P = 0.065$), and a gradient-boosted reweighting is first at four of
them (difference from the linear one $0.059$, $P = 0.21$), so the result does not depend on the
learner. Neither winner takes the other question (the map ranks later vulnerability at $0.29$, the
linear reweighting captures 28.6 per cent): capture rewards knowing where the water goes, the rank
correlation who is fragile when it comes.

The pre-registered blind remedy sits between the two. Averaging the index with the map raises
capture by $0.141$, at nine of eleven origins ($P = 0.06$); the exceptions, 2019 and 2020, leave
test windows of at most five years dominated by the Emilia-Romagna floods of May 2023. But the map
alone captures more at every origin: inside the index, the map is diluted. Past recorded damage
captures as much as the average (difference $0.001$), so any ranking that looks at losses or at
water beats the index; the map needs no loss record. Two caveats bound the reading: the map is the
2021 edition, later than every origin, a look-ahead the 2017 edition removes only at the 2014
split; and test windows overlap, so eleven origins are not eleven independent tests.

\begin{table}[t]
\centering\small
\caption{Rolling-origin forecasts, Italian provinces. Capture: share of damage recorded after the
origin that falls in the top ten provinces, mean over origins 2010--2020. Rank: Spearman correlation
with the revealed trait of later events, mean over origins 2010--2014. Last column: origins, of
eleven, at which the ranking captures more than the published index.}
\label{tab:tournament}
\begin{tabular}{lccc}
\toprule
Ranking & Capture@10 & Rank $\rho$ & Ahead of index \\
\midrule
National flood hazard map & $0.439$ & $0.292$ & 11 \\
Index averaged with the map & $0.299$ & $0.265$ & 9 \\
Components, gradient boosting & $0.298$ & $0.481$ & 9 \\
Recorded damage, previous ten years & $0.298$ & $0.261$ & 10 \\
Components, ridge & $0.286$ & $0.422$ & 11 \\
Population & $0.209$ & $0.116$ & 9 \\
Random & $0.202$ & $0$ & 9 \\
Revealed trait, previous events & $0.197$ & $0.345$ & 9 \\
Published index & $0.159$ & $0.033$ & -- \\
\bottomrule
\end{tabular}
\end{table}

\section{Recorded loss and public transfers}

Over 2013--2022, in 2022 prices, recorded damage across the ninety-three provinces with a figure
amounts to EUR 12.29 billion and state transfers to EUR 1.78 billion, leaving EUR 10.51 billion,
85.5 per cent of the total or EUR 171 per inhabitant. The residual is a bound, not an estimate:
recorded damage is incomplete, which biases it down, while reconstruction programmes, the European
Union Solidarity Fund, tax relief and regional funds are outside the transfer data, which biases it
up.

Coverage is low and uneven. The median province recovers 14.8 per cent of its recorded damage and
sixteen provinces less than five per cent. The elasticity of transfers to recorded damage is
$0.44$, so larger events are proportionally less covered. A national emergency declaration is
predicted by affected population ($t = 4.2$) and not by hazard, exposure, income or the index. The
same holds for the other public responses the data allow to observe: Copernicus rapid-mapping
activations and emergency public works under the procurement register follow people affected
($t = 4.8$ and $2.7$), not recorded euros ($t = 1.4$ and $1.5$). The quantity the index should
predict is not the one the responding institutions track.

Private insurance does not fill the gap; it follows income. Premium density is almost entirely
explained by income per head ($R^2 = 0.84$), and at equal income and loss history it is about 30
per cent lower per index point ($t = -3.8$). The residual loss per inhabitant correlates at $-0.27$
with insurance density. The uninsured margin is thickest where assessed vulnerability is highest,
the premise of the catastrophe insurance obligation introduced for firms by Law 213/2023.

\section{The allocation of prevention expenditure}
\label{sec:prevention}

Prevention is upstream of damage, and the register allows the direct question: where does the
money go? \citet{karim2020} find Bangladeshi sub-district allocations correlated with both flood
hazard and socio-economic vulnerability. Italy's allocation follows neither of the quantities
that predict loss.

Scored as a ranking on the yardstick of Section~\ref{sec:index}, prevention funding per head does
no better than chance. In a pre-registered test on the seventy provinces with every ranking, it
places 6.8 per cent of the damage recorded after 2014 inside its top ten, against 12.6 for a random
draw and 23.0 for the blind index-and-map combination ($P = 0.065$ for the difference). The ten
provinces that took 66 per cent of that damage received 8.5 per cent of the funding while holding
13.2 per cent of the residents. Scored against the damage of 2005--2014 instead, the money again
does no better than chance: it does not chase earlier floods either.

\begin{figure}[t]
\centering
\includegraphics[width=\textwidth]{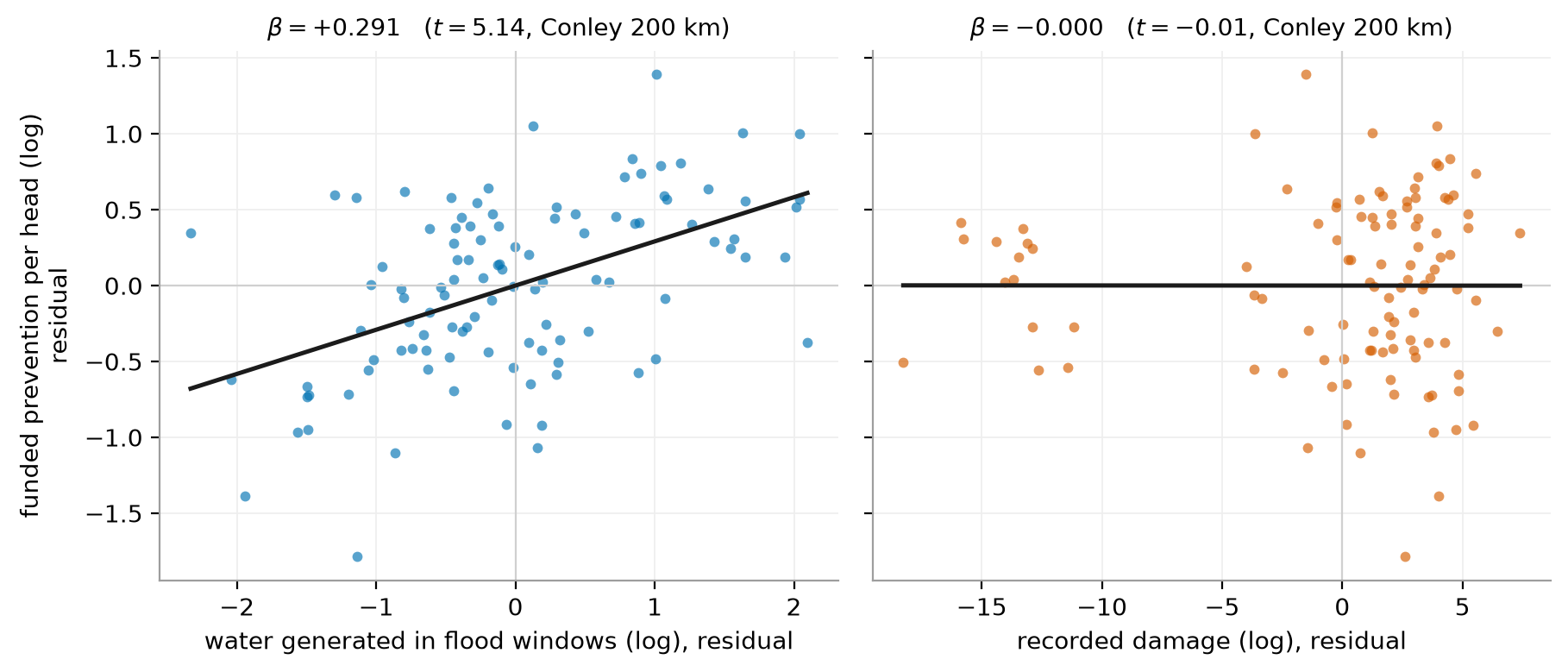}
\caption{Funded prevention per head against water generated within the province (left) and against
recorded damage (right), Italian provinces. Added-variable plots at the specification that controls
for population, income per head and the number of times the province appears as hit, before area
is added; the slope of each line is the coefficient quoted in the text.}
\label{fig:prevention}
\end{figure}

The money follows the water. Regressing log funding per head on the log volume of water generated
within the province during flood windows, with population, income and the number of times the
province was hit held fixed, gives an
elasticity of $0.29$ ($t = 5.14$ under Conley errors). Volume scales with territory, which has a
claim of its own on soil-defence money. With area added, the elasticity falls to $0.190$
($t = 3.20$), and it survives a horse race against the index, recorded damage, external inflow and
both hazard maps ($0.221$, $t = 3.72$). The association is carried by the cohort funded under the
1998 instrument. In the post-2010 cohort, with area fixed, water loses significance and recorded
damage turns weakly positive ($0.019$, $t = 2.03$ under HC1). Over the whole register, before area is added,
recorded damage returns $-0.000$ ($t = -0.01$), and with area fixed the flood hazard map that
predicts loss in Section~\ref{sec:index} carries $0.024$ ($t = 0.39$). Figure~\ref{fig:prevention}
shows the water and damage coefficients before area is added.

Where the water is generated is also where the money stops. The share of committed funds that has
never entered execution rises with generated volume ($0.040$, $t = 4.73$ under Conley errors;
$0.039$, $t = 3.39$, with area fixed). Upstream provinces are both the target of the programme and
the point at which it fails to become construction.

None of these associations is causal. The dated projects of \citet{ricciotti2024} allow a province-year panel
with province and year fixed effects: after a recorded flood or landslide, neither the count nor
the funding of new prevention projects rises in the following three years. What rises is the
number of smaller hydrogeological works on treasury and cohesion lines, by 23 per cent in the year
of the event ($z = 5.4$), with no anticipation and no matching rise in funding. After a disaster
the pipeline reacts with repair-sized works; the prevention programme does not re-target.

The dated projects also allow one causal question: does completing a flood-defence work reduce
what the municipality later needs in emergency? In a pre-registered stacked difference-in-differences,
882 municipalities whose first flood-defence work was completed in 2016--2021 are compared with
2{,}122 funded municipalities whose works finished later or not at all, within the same province and
year, so that weather and regional administration are held fixed. The outcome is whether any
emergency public works are procured in the municipality in a year, which happens in 20 per cent of
treated municipality-years before completion. The years before completion show no differential trend
(joint $p = 0.44$). In the three years after it the probability changes by $-1.1$ points (standard
error $1.6$, one-sided $p = 0.24$), with a 90 per cent interval of $[-3.7, 1.4]$, and no larger
effect in wet years. At municipal and annual resolution, completed prevention leaves no detectable
trace in emergency spending. This bounds the effect, excluding reductions larger than about a fifth
of the baseline, and is not evidence that prevention fails.

\section{The structure of joint losses}

\subsection{Provinces are hit in groups}

Co-occurrence read off the projection of the event-by-province incidence matrix is mostly
generated by how large and how often hit each province is. Against the bipartite configuration
model \citep{saracco2017}, which holds both fixed, only 131 of 3{,}154 co-occurring Italian pairs
survive at a five per cent false discovery rate \citep{benjamini1995}, so statistics on the raw
projection are not reported.

Two properties of the incidence survive the same null. It is less nested than chance
\citep[NODF $12.54$ against $15.83$, $z = -5.24$;][]{almeidaneto2008}, in Italy and in Europe:
joint incidence follows no single ordering of provinces. And it is organised in groups. Of the
province triples whose three pairs each co-occur somewhere, 88.0 per cent appear together in a
single event, against 67.1 per cent under the null ($z = 8.2$), and 89.3 against 56.5 in Europe
\citep{benson2018,landry2024}. The excess grows with group size, to a ratio of $2.78$ at order
five.

\begin{figure}[t]
\centering
\includegraphics[width=\textwidth]{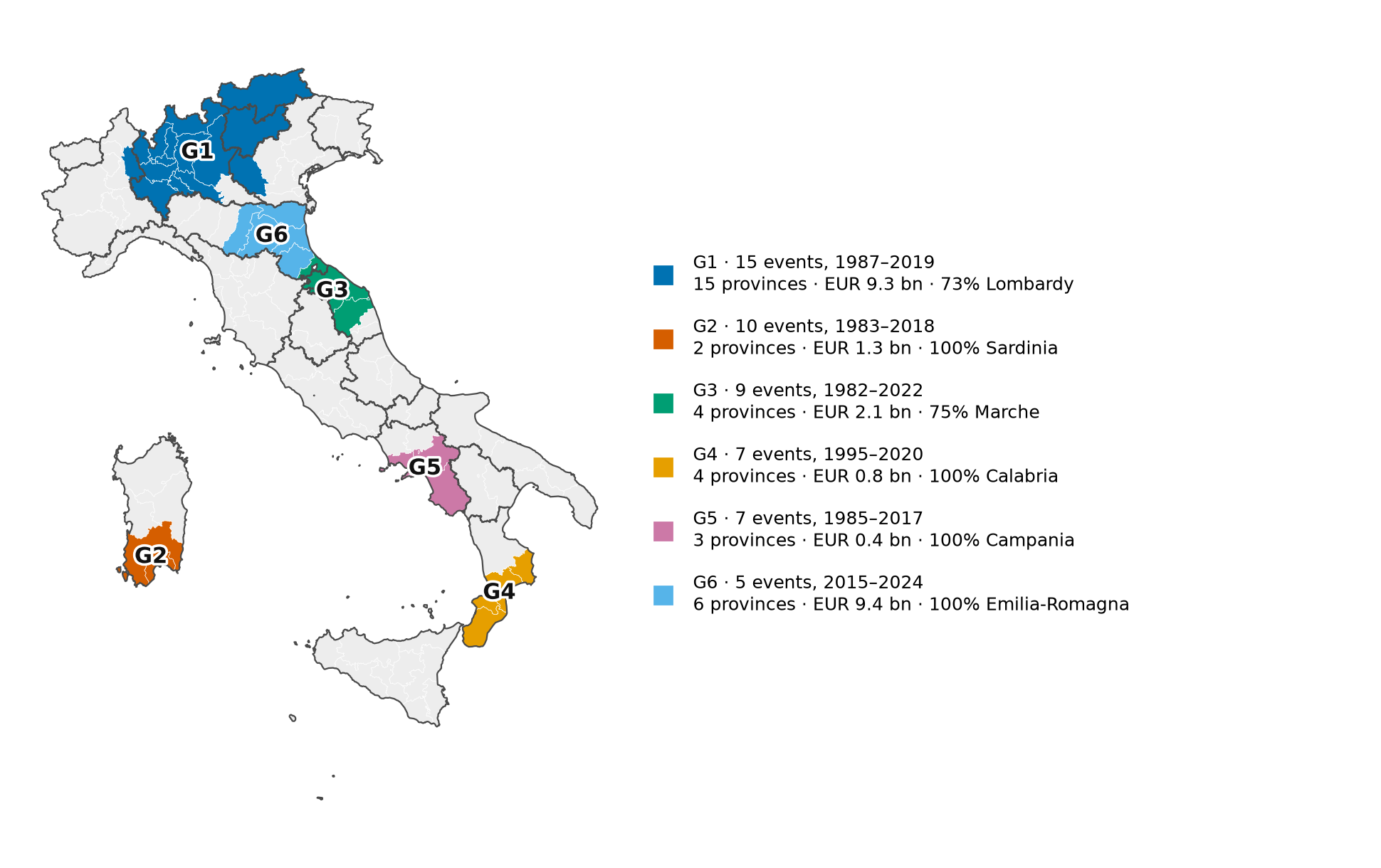}
\caption{Cores of the six most recurrent joint-loss groups, ties broken by recorded loss, of the
twenty-eight found at a Jaccard threshold of one half. Grey provinces belong to no selected core.
Dark lines are regional boundaries. Rimini belongs to two selected cores and is drawn in the one
where it appears more often.}
\label{fig:groups}
\end{figure}

Linking events whose province sets overlap at a Jaccard coefficient of one half, and taking as the
core of each family the provinces present in at least half of its events, Italy has twenty-eight
recurrent groups covering 108 of 175 events and 68.5 per cent of recorded damage, against 8.9 groups
and 13 per cent of events under the null. The observed configuration lies outside the ensemble at
every threshold from $0.40$ to $0.75$. Cores sit inside a single river basin at a median of 100 per
cent and inside a single region at a median of 84 per cent, and they span decades. They also
return after being found: built on events up to 2014, they match 17.9 per cent of the
multi-province events of 2015--2024 against none under the null, and 21 per cent in Europe. Nor are
they an artefact of the register. In 83 European floods mapped from orbit \citep{tellman2021} the
same excess appears ($z$ between 7 and 16), and among units with water on the ground, whether a
loss is recorded depends on the flooded area and not on income per head ($-0.03$, interval
$[-0.12, 0.06]$). Figure~\ref{fig:groups} maps the six most recurrent groups. The largest runs from
Piedmont across Lombardy to the eastern Alps, and the costliest is the eastern Emilia-Romagna set.
Neither is a region.

\subsection{The index errs by group}

Assigning each province to its strongest group, 56.6 per cent of the variance of the index's
ranking error lies between groups, against 26.1 per cent when the labels are permuted
($p = 0.0001$). Because the error follows a North--South gradient that any compact partition would
pick up, the sharper benchmark is random compact partitions of the same sizes, which place 43.2 per
cent between pieces (90 per cent envelope $[32.8, 53.4]$); the groups still exceed it ($p = 0.012$).
Geography accounts for about half of the excess, the groups for the rest. Both European samples
place 54.1 per cent between groups, against 42.5 and 40.3 per cent under compact partitions
($p < 0.001$). Residualising the error on how often a province is observed changes none of this.

\subsection{The administrative unit is not the unit of the problem}

Of the basin-to-basin water deliveries measured in flood windows, 14.3 per cent stay inside one
province and 62.6 per cent cross a regional border. Twenty-five of the 107 provinces receive most
of their water from outside their region. That geography reaches the outcomes. Revealed
vulnerability rises with the externally generated share of arriving water ($t = 2.22$ with income
and size), while public transfers per head fall with it ($t = -2.15$). Prevention money is decided
by region and national programme, transfers by declaration, and declarations are regional by
procedure: 119 of the 123 national emergencies cover a single region. Within an event, loss also co-moves along
the regional line: in a spatial Durbin model with contiguity, water delivery and region as
competing neighbour definitions, only the regional term survives ($0.35$, $t = 3.50$; water $0.24$,
$t = 0.94$). This may reflect loss assessed and managed region by region; the design cannot
separate that from shared exposure.

\subsection{Risk sharing across the boundary}

If losses arrive in groups that sit mostly inside one region, a region is a poor unit in which to
pool them. Pools with the sizes of the Italian regions, drawn on events up to 2014 and scored on
the damage of 2015--2024, lose on average 78.8 per cent of their damage to their single worst
event, against 70.2 per cent for compact random pools of the same sizes, 63.8 for scattered ones
and 44.7 for one national pool. Regions are the worst of the designs compared.

In Europe the pooling unit is the country. Resampling the observed footprints of events up to 2014
and summing the reserve each pool needs for a 1-in-50-year annual loss, national pools require 2.29
times the reserve of a single European pool, and scattered cross-border pools of the same sizes
1.64 times (90 per cent band $[1.54, 1.76]$). National pooling thus needs 39 per cent more reserve
for the same cover, 38 per cent with Italy excluded. This updates, on recorded subnational losses,
the solvency cost of national pooling estimated by \citet{prettenthaler2017}. Scattered pools are a
benchmark, not a proposal. Dependence matters too: treating provinces as independent understates
the Italian 1-in-50 loss by about a quarter, while a model that matches every co-hit pair
reproduces it. It reproduces the tail, not the structure: the same model predicts 370 province
triples per event against 630 observed and 11{,}115 five-province sets against 38{,}809, and
misplaces the size of events. Pairs carry the national reserve; the groups are higher-order.

The groups do not, however, improve allocation. In a pre-registered test, ranking provinces by the
past losses of their group captures no more of the 2015--2024 damage than their own past losses
($P = 0.38$), in Italy or in Europe. Five Emilia-Romagna provinces hit in 2023 carry 47 per cent of
that decade's damage. The structure tells how losses should be shared, not yet where they will
fall. An exploratory percolation exercise on the European units open to exploration points the
same way. Protecting 30 per cent of units by hypergraph eigenvector centrality breaks the co-loss
network most (largest connected component 10 per cent of units, against 18 at random and 26 under
the published index), while protecting by own past loss removes the most loss (44 per cent left,
against 70 at random and 74 under the index). Breaking connectivity and removing loss call for
different rankings, and the index serves neither.

\section{The same storms in a warmer climate}

Scaling each event's precipitation and re-ranking it against the unchanged 1950--2025 climatology
of the province it hit, with exposure, insurance and behaviour fixed, brackets how much more
extreme the same storms become. The Clausius--Clapeyron rate of seven per cent per degree raises
the share of province-events above their local 95th percentile from 47.3 to 63.5 per cent at
$+2$ degrees. Ten downscaled CMIP6 models \citep{thrasher2022}, compared between 1985--2014 and
2041--2070, give far less for the multi-day totals behind these floods ($+1.6$ per cent per degree
for the five-day maximum, $+4$ for the one-day maximum) and raise the share to 53.8 per cent under
SSP2-4.5 and 54.7 under SSP5-8.5. The uniform rate is an upper bound, and neither scaling resolves
hourly convective extremes, which can exceed it \citep{lenderink2008}. Modelled loss rises by 0.7
to 5.8 per cent under the models, and its interval under the uniform rate contains zero, so the
damage side is not reported as a result. What survives is the geography: the percentage increase
in modelled loss correlates
negatively with the published index ($-0.21$ to $-0.31$, same sign in nine of ten models) and
positively with insurance premiums per head ($+0.22$ to $+0.27$). The additional hazard falls
where the index reports safety and cover is thickest, in the North.

\section{Limitations}

The central claims are predictive, which is what an instrument that allocates money must deliver:
the index is a forecast and is judged by what follows. The associations behind them are not
causal. Prevention spending is endogenous to need, transfers follow declarations that follow
damage, and the recurrent groups are estimated from the same events on which outcomes are
measured. The one causal estimate, on completed prevention works (Section~\ref{sec:prevention}),
is a pre-registered null with a bound.

The damage register under-reports. Satellite flood extents show that 22.5 per cent of
province-event cells with detected water carry a record, against 2.1 per cent without: the register
sees water but misses about three quarters of the cells where it is detected. Emergency public
works rise with runoff in province-windows where no loss was recorded ($t = 4.25$). Every monetary
figure here is a floor.

Only hydro-meteorological events are studied: the register records other perils one province at
a time, leaving a within-event design nothing to compare, and landslides carry no economic figure.
Half of the twenty-eight groups rest on two events. They carry 12.8 per cent of the damage
attributed to groups, which is why the groups mapped are selected on recurrence.

\section{Conclusion}

The index the European Union publishes to compare territories before a disaster does not order
Italian provinces by the damage they take. It is not short of information; it weights the wrong
information. It measures development, while losses follow water, arriving in recurrent groups of
provinces that share a river basin and carry two thirds of recorded damage. Reweighting its
components without losses does not help. Two sources already in public hands do, each for one
question. For where damage will land, the national flood hazard map captures more of later damage
than the index at every forecast origin from 2010 to 2020, alone better than averaged with it. For
who loses more from the same water, the index's own components, reweighted on past losses, rank
provinces where the published weights do not.

The loss the ranking misses is paid locally. Over a decade, EUR 10.51 billion of recorded damage
was not met by national transfers, concentrated where private insurance is thinnest. Prevention
funding of EUR 21.50 billion follows territory and the generation of water, lands on the provinces
that later lost no better than chance, and almost half of it has never become a building site.

Two implications follow. Assessment instruments meant to guide allocation should be validated
against realised outcomes at the scale at which they are published, with blind reweightings in the
comparison, and scored separately on where losses land and on who loses more, since no single
ranking, and no average of two, served both here. And because
losses are correlated along the river network and arrive in groups that sit inside one region or
one country, funds that pool them should
cross those borders: regions are the worst pooling unit in Italy, and national pools in Europe need
38 to 39 per cent more reserve than cross-border pools of the same sizes.

\vspace{1em}
\footnotesize
\noindent\textbf{Data and code.} All datasets used are public. Analysis code and derived tables are
available from the author. \textbf{Declarations.} Sole authorship, no funding, no competing
interests. Generative AI tools were used for code drafting and language editing; all results,
statistics and interpretations were produced and verified by the author.

\normalsize
\bibliography{refs}

\end{document}